# Subclustering and Luminous-Dark Matter Segregation in Galaxy Clusters

**A. Serna, J.-M. Alimi**

*Laboratoire d'Astrophysique Extragalactique et de Cosmologie.*
*Observatoire de Paris-Meudon. 92195 Meudon, France*

and

**H. Scholl**

*Département Cassini. Observatoire de la Côte d'Azur. BP 229,*
*06304 Nice, France*

# Abstract

We have performed a series of high resolution N-body experiments on a Connection Machine CM-5 in order to simulate the formation of galaxy clusters gravitationally dominated by a massive dark background. In accordance with previous authors we find an extremely inhomogeneous evolution where subcondensations are continually formed and merged. Final distributions do not present subclustering and galaxies are more centrally condensed than dark matter particles. We have found that such a luminous-dark matter segregation persists even when inelastic encounters of dark halos around galaxies are taken into account.

We show rigorously that such a segregation is due to dynamical friction within the subclusters, while an interpretation of this effect as a biasing process in initial conditions, is ruled out. We are then interested in the origin of subclustering. We find that galaxy-dark matter segregation is mainly due to the substructures formed from local gravitational perturbations produced by the presence itself of massive bodies (galaxies), while the subclusters formed from initial Poisson density fluctuations. Consequently, the final amount of segregation depends only very slightly on the initial fluctuation spectrum.

Physical parameters characterizing the model of protocluster have, however, an important influence on the final result. This dependence has been qualitatively explained by means of a simple dynamical model and has been numerically analyzed. We find that final distributions without segregation are only possible for extreme values of these physical parameters.

We then conclude that such a segregation effect cannot be in general avoided in any hierarchical clustering scenario. Consequently, all observational analyses assuming that galaxies in clusters trace mass are misleading and their conclusions have to be reconsidered.

**Subject headings:** cosmology – dark matter – galaxies: cluster of – galaxies: clustering

# 1. Introduction.

There exists a general agreement that self-gravitating systems go through two different stages in their evolution: first, a fast phase of violent relaxation, where the dynamics is controlled by a collective potential, and then a slow two-body relaxation phase, where the dominant role corresponds to binary collisions. The final distribution of energy per unit mass which results from violent relaxation is independent of particle masses (Lynden-Bell 1967), so one expects that no mass segregation will be produced during this phase. On the other hand, the estimates of $M/L$ for rich galaxy clusters suggest that a large part of their mass may well be in some dark form. Current theories of galaxy formation admit a wide range of possible candidates for this dark matter. Within the context of hierarchical clustering theories, objects in the range of $10^{-5} ev$ to $10^6 M_\odot$ seem possible (Lacey & Ostriker 1985; Turner 1987; Gnedin & Ostriker 1992; Wambsganss & Pacynski 1992).

N-body simulations show that formation of clusters containing both galaxies and dark matter can be however rather complicated. Numerical calculations by White (1976), Evrard (1987) and West & Richstone (1988, hereafter WR) imply that clusters of galaxies have an extremely inhomogeneous evolution where subcondensations are continually formed and merged, giving rise to final distributions with a considerable spatial segregation between the two mass components. This result can be also inferred from several other works (e.g., Roos & Aarseth 1982; Barnes 1984; Cavaliere et al. 1986). A significant luminous-dark matter segregation in galaxy clusters would imply that virial mass estimates based on galaxy coordinates have a great error (Limber 1959; Aarseth & Saslaw 1972; Smith et al. 1979; Smith 1980, 1984; Bailey 1982; Hoffman, Shaham & Shaviv 1982; Barnes 1983, 1984; Valtonen et al. 1985; Navarro, Garcia-Lambas & Sérsic 1986; Cowie, Henriksen & Mushotzky 1987; Evrard 1987; Merrit 1987; The & White 1986, 1988). WR pointed out that the density parameter values $\Omega \simeq 0.1$–$0.2$ obtained from galaxy clusters could in that case be consistent with true values of $\Omega = 1$.

A possible mechanism leading to that segregation could be the exchange of energy from galaxies to dark matter particles produced by dynamical friction inside the subclusters. By using the standard formula (Chandrasekhar 1943) to estimate the dynamical friction time within a subcluster, West and Richstone (WR, West 1990, Richstone 1990) found

$$T_{df} = \frac{0.1 N_g T_{cr}}{x(1-x)(1+f)\ln\Lambda} \quad (1)$$

where all variables refer to a subcluster, $N_g$ being the number of galaxies in the subcluster, $x$ the galaxy mass fraction, $f$ the ratio of dark to luminous particle mass, $\ln\Lambda$ the usual Coulomb logarithm and $T_{cr}$ the crossing time,

$$T_{cr} = R/v \quad (2)$$

with $R$ and $v$ being the subcluster radius and galaxy velocity, respectively. For $N_g = 1$, $x = 0.1$, $\ln\Lambda = 3$ and $f \simeq 0$, equation (1) implies $T_{df} \lesssim T_{cr}$. So, the friction time in substructures could be rather short and galaxies would go to the lump centers on a few crossing times.

However, there are still some questions referring to this luminous-dark matter segregation which are sufficiently important and controversial to be examined in detail. To this end, we have performed a series of N-body simulations where the simplest cosmological picture and assumptions are taken in order to establish a well-posed dynamical problem. Section 2 discusses the initial conditions and the N-body code used for these simulations.

First, although the above interpretation given by WR is reasonable, it has not been rigorously proved. In principle, other possible explanations, as some "biasing" effects resulting from the initial distribution of galaxies in the deepest potential wells are also considered to explain that segregation (White 1993). Both possibilities will be tested (Section 3) from a convenient series of numerical experiences.

Another question which still holds without a fully satisfactory solution is that concerning the development of subclustering during the system expansion and collapse phases. The answer to this question has a fundamental importance because it could be related to some possibility of avoiding this effect. WR claimed that subclusters leading to a final segregation are formed from small-scale statistical fluctuations on the initial particle distribution. Different initial density fluctuation spectra would then imply some quantitative differences in the final amount of segregation. The extreme case where small-scale Poisson fluctuations are initially suppressed could then imply similar resulting distributions for both mass components. A grid simulation performed by these authors seems to support that, but their conclusions are not clear because in the same numerical experiment they suppressed the cluster expansion phase, which could be a serious default in their analysis. A second type of subclusters, whose influence on the final results has not been analyzed, are those formed from the local gravitational perturbation produced by the presence itself of massive objects (galaxies) in a background of much lighter and more numerous particles (dark matter). If such a subclustering mechanism is important, lumps of dark matter would be always formed around galaxies (provided clusters form in a bottom up manner) and segregation would develop even from extreme initial conditions as those of a grid simulation. Section 4 is devoted to analyze this subject.

Finally, we must also analyze the influence of the different physical parameters characterizing the model of protocluster on the final amount of segregation. Equation (1) gives some information about the physical features which could modify the final results. However, the parameters appearing in that equation are not known a priori because they refer to subclusters while calculations start by specifying the overall cluster characteristics. A simple infall model is developed in Section 5 to estimate how the initial cluster parameters enter in equation (1). This will allow us to interpret qualitatively a sequence of simulations where the influence of different physical parameters is analyzed. Infall models for the formation of condensed objects were first constructed by Gunn & Gott (1972) and extended later by Gott (1975), Gunn (1977), Fillmore & Goldreich (1984), Bertschinger (1985) and White & Zaritsky (1992) for a wider variety of conditions.

Inherent in all the above simulations is an assumption whose possible influence must be tested: dark matter is initially distributed as a smooth background and galaxies interact elastically. However, it has been suggested (Eyles et al 1991) that segregation could be avoided by taking dark matter particles initially distributed as extended halos around galaxies. Inelastic encounters between these galaxy-halo systems could lead to loss of energy by the stripped dark material. In this case, non-luminous particles would progressively concentrate in the core as they are removed from galaxies. In order to analyze this possibility, we have simulated (Section 6) the dynamical evolution of a cluster of galaxies containing massive dark halos.

## 2. Method.

### 2.1 Cluster Models.

As in WR, we consider the simplest cosmological scenario where a cluster of galaxies is formed from the spherical collapse of a primordial density fluctuation. This overdense region initially expands with the Hubble flow, eventually reaches a maximum radius and then turns around and begins to contract.

In order to simulate this simple model, we have considered a self-gravitating system containing both heavy (galaxies) and light (dark matter) softened particles of individual masses $m_g$ and $m_{dm}$, respectively. Initial particle positions were typically assigned at random within a uniform sphere of radius $R_{in}$. The initial velocity field was a Hubble flow, $\mathbf{v}_i = H_{in} \, \mathbf{r}_i$, where $\mathbf{r}_i$ is the position vector of each particle i and $H_{in}$ is the Hubble parameter at the starting time. For such an expanding uniform sphere, initial kinetic and potential energies are given by $K_{in} = \frac{3}{10} M_T H_{in}^2 R_{in}^2$ and $U_{in} = \frac{3}{5} G M_T^2 / R_{in}$, respectively, where $M_T$ is the total mass. The Hubble parameter is then related to the maximum radius, $R_m$, by

$$H_{in} = \sqrt{\frac{2GM_T}{R_{in}^3}\left(1 - \frac{R_{in}}{R_m}\right)} \qquad (3)$$

and the time required for expansion and subsequent collapse is

$$t_{coll} = 2\pi\sqrt{\frac{R_m^3}{8GM_T}} \qquad (4)$$

Throughout this paper we have used units in which $G = M_T = 1$ and $R_{in} = 1/3$. Previous works (White 1976, Cavaliere et al. 1986, WR) have chosen $H_{in}$ in order to obtain $R_m/R_{in} \simeq 6$. When comparing with clusters such as Coma, this expansion factor gives a collapse time of the order of $10^{10}$ yr and an initial redshift of about 25 in an Einstein-de Sitter Universe. The expansion factor is also related to the initial values of the density parameter $\Omega_i$ and the density contrast $\delta_i$ (Peebles 1980). Clusters different from Coma or models with a shorter or larger post-collapse evolution might be better described by using a somewhat different $R_m/R_{in}$ value. In the present study, we have taken $R_m/R_{in} \simeq 6$ as the standard expansion factor value, but we have also analyzed the effects of considering smaller or larger values (see below section 5).

The numbers and masses of the two types of particles were chosen in such a way that while galaxies are individually much more massive, the total mass of the cluster is dominated by the non-luminous component. In most of our simulations, $2^{13}$ particles were considered, 30 of which were taken as galaxies. The remaining bodies represented dark matter particles with a mass $m_{dm} \simeq m_g/30$ in order to have a total non-luminous mass fraction of $\eta = 0.9$. In real clusters, the ratio $m_g/m_{dm}$ must be many orders of magnitude higher. Fortunately, tidal effects due to individual dark matter particles seem to be unimportant and $m_{dm}$ enters as $(1 + m_{dm}/m_g)$ in Chandrasekhar's formula for dynamical friction. One then expects that the effects analyzed in this paper will be only very slightly modified by the technical limitations of the dark matter mass assignation.

Most of our present simulations do not incorporate galactic collision effects such as tidal stripping and merging. Despite the efforts of a number of workers (Richstone 1975, 1976; Roos & Norman 1979; Dekel, Lecar & Shaham 1980; Merrit 1983, 1984; Aguilar & White 1985), the relevant cross-sections for inelastic collisions are still not well known. By including these effects, one may well commit a similar error as by neglecting them. However, these processes are clearly important and their possible influence on the final results must be tested. To this end, we have also performed simulations where the inelasticity of halo encounters is taken into account. The two-component approach used for that simulation is similar to that previously considered by Evrard (1986).

Table 1 presents a summary of our N-body calculations.

### 2.2 N-Body Code.

Particles of mass $m_i$ and $m_j$ interact through the softened potential:

$$\phi_{ij} = Gm_i m_j / (r_{ij}^2 + \epsilon_i^2 + \epsilon_j^2)^{1/2} \qquad (5)$$

where $r_{ij}$ is their separation. The softening parameter $\epsilon_i$ is essentially a particle radius. We have adopted the usual convention $\epsilon_i = \epsilon_0 m_i^{1/3}$ where particles are taken as Plummer spheres of identical central density (Farouki & Salpeter 1982; Farouki, Hoffman & Salpeter 1983). In our simulations, $\epsilon_0$ was chosen in order to obtain $\epsilon_i = 0.025$ for the heaviest particles.

The most obvious algorithm to integrate the equations (5) is direct summation (Aarseth 1972, 1985). Unfortunately, the number of operations in this algorithm grows as $N^2$. On classical sequential computers it is at present not possible to consider a large number of particles ($N \leq 10000$) and to perform many numerical simulations. On the other hand, the direct summation algorithm is specially adapted for massively parallel computers like a Connection Machine. In this paper, most of numerical simulations have been performed on a Connection Machine CM-5 5 by using a "Digital Orrery" algorithm, first developed on a Connection Machine CM-2, (Hillis & Barnes 1987, Alimi & Scholl 1993). In this case, one physical or virtual processor is assigned to each particle. One can then imagine two rings of processors containing both the same set of N-body coordinates. One ring starts to turn around stepwise by the use of CM instructions. At each step, forces between all adjacent couples of bodies placed on the two different rings are calculated in parallel. A complete force calculation between all particles requires $N - 1$ such steps. To illustrate the computational requirements, for a simulation with 16384 particles, the main part, computing all interacting forces between all particles, took less than 20 s per timestep. The accuracy of the integration can be illustrated by the variation of the total energy throughout a complete calculation ($2t_{coll}$). The energy violation in a 16384 particle system was smaller than 0.001 per cent in a typical timestep, and 0.01 per cent for the overall simulation.

## 3. Origin of the Luminous-Dark Matter Segregation.

The time evolution of a typical numerical simulation is shown in Figures 1. The small points represent dark matter particles and the large points represent galaxies. Like previous authors (White 1976; Roos & Aarseth 1982; Barnes 1984; Cavaliere et al. 1986; Evrard 1987; WR), we find that during the expansion phase, several substructures appear in the system. These substructures merge during the phase of collapse giving rise to a final configuration without considerable subclustering and where galaxies are more concentrated than dark matter particles.

In order to analyze the time evolution of this galaxy-dark matter segregation we have plotted (Figure 2) the harmonic mean separation

$$R_{hm} = \frac{N(N-1)}{2} \left( \sum_i \sum_{j<i} \frac{1}{r_{ij}} \right)^{-1} \tag{6}$$

for both mass components. A segregation between the galaxies and dark matter develops rapidly on a timescale much shorter than the cluster collapse time (spherical top-hat model). As we have quoted in the introduction, there are two possible mechanisms which could explain such a segregation: dynamical friction in subclusters and biasing effects in the initial conditions. In order to elucidate the viability of these two mechanisms, we have performed a series of simulations where the effects of both processes have been separated.

To this end, we repeat the previous simulation (A1) but this time with only dark matter particles (Figure 3). In a first experience (B1), those particles placed in the deepest Poisson potential wells are called "galaxies". Since the individual mass of these "galaxies" is equal to that of dark matter particles, we cancel in this way any possibility of dynamical friction. Figure 4a shows the time evolution of $R_{hm}$ for both species of particles with only a very small final segregation. This is compatible with no final segregation if we take into account the uncertainties.

In a second experience (B2), particles placed in the deepest potential wells of the previous simulation are again called "galaxies". However, we now change the mass of these "galaxies" and suppress a convenient number of surrounding dark matter particles in order to get a ratio $m_g/m_{dm}$ similar to that of simulations A1. The suppression of surrounding light particles ensures that the mass distribution is not changed on scales larger than those of massive particles. We have in this way the same initial statistical fluctuations and the same biasing as in the previous experiment. Any difference between the subsequent evolution and that observed in the first experiment must be then attributed to dynamical friction. The evolution of the harmonic mean separation shows now an important galaxy-dark matter segregation (Figure 4b). We then conclude that only dynamical friction in the subclusters seems a viable explanation for such a segregation.

## 4. The Origin of Subclustering.

The interpretation of this kind of segregation as a consequence of dynamical friction within the substructures implies that it must depend strongly on the development of subclustering during the cluster evolution. WR consider that subclusters form from statistical fluctuations on the initial particle distribution. Consequently, the amount of segregation which develops on different scales in a hierarchical clustering scenario must depend somewhat on the initial density fluctuation spectrum. In order to test this, they simulated the collapse phase of a cluster where particles were placed at the points of a cubic grid in order to suppress initial small-scale Poisson fluctuations. No significant segregation was then observed to occur between galaxies and dark matter, so they concluded that their result was due to the absence of initial fluctuations.

Some number of subclusters clearly appears from initial Poisson fluctuations. The comparison between figures 1 and 3, where the only difference is the mass assignation for the 30 particles taken as galaxies, shows however some differences on the initial subcluster positions. The origin of the overdense regions in Figure 3 is in fact purely statistical and several of them can also be identified in Figure 1 at the earlier evolutionary stages. However, there also exist on Figure 1 several subclusters which are always related to at least one galaxy. These last regions must be interpreted as the result of the local gravitational perturbation produced by the presence itself of very massive objects (galaxies). We must analyze what is the importance of each kind of subclusters on the final segregation in order to test if this effect is avoidable and to test its possible dependence on the initial density fluctuation spectrum.

To this end, we have performed a first experience (C1), where the expansion phase is also canceled but now the particle positions are assigned at random within a sphere of radius $R_{in} = 2$, so our system has initial statistical fluctuations. The collapse phase of this system for a time roughly equal to $t_{coll}$ leads to a final result which is very similar to that obtained by WR in their grid simulation (Figure 7 of WR), that is, no significant segregation is found (Figure 5). This result indicates that the expansion phase plays a fundamental role in order to obtain final segregated distributions. Dynamical friction is only effective within small subclusters containing a low number of galaxies. The lifetime of this kind of substructures during the collapse phase is too short because they rapidly merge giving rise to a great central overdensity. If the cluster reaches its maximum radius without developing previously such subclusters, there is no segregation because dynamical friction has not enough time to play a role inside them. Consequently, the grid simulation performed by WR does not prove a statistical origin of subclusters leading to a final galaxy-dark matter segregation.

In two last numerical experiences (C2 and C3), we have simulated both the expansion and collapse phases beginning from initial particle distributions where small-scale Poisson fluctuations were suppressed. In simulation C2, all particles were initially placed on the same grid while in C3 a different grid was used for each mass component. Figure 6 shows the time evolution of this last experience. In both cases, we find that subclusters are always associated to galaxies. Consequently, the origin of these substructures is essentially due to the local gravitational attraction of galaxies. We obtain final segregations (Figures 7) which are similar to that obtained in simulation A1 from initial random positions (see Figure 2). This suggests that segregation is mainly related to such kind of subclusters, instead of those formed from initial statistical density fluctuations. It is interesting to note from Figures 6 and 7b that the extreme symmetry of initial conditions in C3 leads to an evolution where all mergers are produced simultaneously at $t \simeq 0.8 t_{coll}$. The $R_{hm}$ for the two mass components also shows that spatial segregation appears suddenly at this instant. The segregation is then latent in subclusters, but it is only communicated to the overall cluster after merging, when redistribution of particles according to their energies can take place.

## 5. Dependence on Physical Parameters.

In the preceding sections we have proved that the final luminous-dark matter segregation depends only very slightly on the initial fluctuation spectrum. However, dynamical friction within subclusters can be more or less effective for different choices of physical parameters characterizing the initial protocluster. Since the subclusters which mainly produce this effect are those formed from the local gravitational attraction of galaxies, we can estimate the way in which these parameters enter in equation (1). To this end, we have developed a simple infall model of formation and evolution of subclusters and we have then performed a sequence of simulations which will be qualitatively interpreted in the framework of that model.

We study the movement of a spherical shell of dark matter around a point galaxy of mass $m_g$. At the starting time, $t_i$, this shell has a radius $\mathbf{r}_i$ and expands with a relative velocity $\mathbf{v}_i = H_{in}\mathbf{r}_i$. We consider that the initial total density within this shell, $\bar{\rho}_i$, is somewhat greater than the critical density $\rho_{ci} = 3H_{in}^2/(8\pi G)$. Then, it will fall on the galaxy after reaching some maximum radius. The collapse time of this shell is given by (Gunn & Gott 1972):

$$t_c = \frac{\pi}{H_{in}} \frac{\rho_{ci}^{3/2}}{(\bar{\rho}_i - \rho_{ci})^{3/2}} \qquad (7)$$

In our case, the initial total density within $\mathbf{r}_i$ is,

$$\bar{\rho}_i = \rho_{DM} + \rho_{1g}\frac{R_{in}^3}{r_i^3} \qquad (8)$$

where $\rho_{DM} = \eta M_T/(\frac{4}{3}\pi R_{in}^3)$ and $\rho_{1g} = m_g/(\frac{4}{3}\pi R_{in}^3)$, $M_T$ and $R_{in}$ being the initial values for mass and radius of the overall cluster.

By substituting equation (8) into (7) and solving for $r_i$, we obtain the initial radius of that shell reaching the galaxy at a given time $t_c$,

$$\left(\frac{r_i}{R_{in}}\right)^3 = \frac{\rho_{1g}\tau^{2/3}}{\rho_{ci} - (\rho_{DM} - \rho_{ci})\tau^{2/3}} \qquad (9)$$

where $\tau \equiv H_{in}t_c/\pi$.

The turn-around radius for a shell of initial radius $r_i$ is (Gunn & Gott 1972):

$$r_m = r_i \frac{\bar{\rho}_i}{\bar{\rho}_i - \rho_{ci}} \qquad (10)$$

In order to estimate the size of a subcluster at a given time $t$, we consider that it is roughly equal to the virial radius $r_m/2$, where $r_m$ is the maximum radius of that shell with a collapse time just equal to $t$. By combining equations (7)–(10), the time dependence of the radius $R_p$ and mass $M_p$ of a subcluster is

$$\left(\frac{R_p}{R_{in}}\right)^3 = \frac{1}{8}\frac{\rho_{1g}\tau^{2/3}(1+\tau^{2/3})^3}{\rho_{ci} - (\rho_{DM} - \rho_{ci})\tau^{2/3}}$$

$$M_p = \frac{m_g\rho_{ci}(1+\tau^{2/3})}{\rho_{ci} - (\rho_{DM} - \rho_{ci})\tau^{2/3}}$$

(11)

where all parameters (except $\tau$) are referred to the starting time.

Note that both $R_p^3$ and $M_p$ are proportional to the individual galaxy mass $m_g$. The density in subclusters is then independent on $m_g$.

Using equation (11) to computate the galactic mass fraction, $x \equiv m_g/M_p$, and the crossing-time (eq. [2]) in a substructure, equation (1) for dynamical friction time gives

$$T_{df} = \frac{0.05 N_g \rho_{ci}^{3/2}(1+\tau^{2/3})^3}{\sqrt{G}\rho_{DM}\tau^{1/3}(\rho_{ci} - (\rho_{DM} - \rho_{ci})\tau^{2/3})(1+f)\log\Lambda} \quad (12)$$

with

$$\Lambda = \frac{0.4 N_g(1+\tau^{2/3})}{(\rho_{ci} - (\rho_{DM} - \rho_{ci})\tau^{2/3})(1+f)} \quad (13)$$

where we have generalized for subclusters containing $N_g$ galaxies. Note that $\rho_{DM}$ is related to the total non-luminous mass fraction $\eta$ and that the initial Hubble parameter $H_{in}$ enters in equation (12) through $\tau$ and $\rho_{ci}$. The $H_{in}$ value is related to the cluster expansion factor, $R_m/R_{in}$, through equation (3).

The Coulomb logarithm in a subcluster (eq. [13]) has been estimated by taking the standard values $b_{min} \simeq G(m_g + m_{dm})/v^2$ and $b_{max} \simeq R_p$ for the minimum and maximum impact parameters (Cohen, Spitzer & Routly 1950, Farouki & Salpeter 1982, Binney & Tremaine 1987), with $v^2 \simeq 0.4 GM_p/R_p$ being the mean square velocity of particles.

Figures 8 display the predicted $(t_{coll}/T_{df})$ values at different evolutionary stages ($t_{coll}/4$, $t_{coll}/2$ and $3t_{coll}/4$) versus the overall mass fraction of dark matter $\eta$, the expansion factor $R_m/R_{in}$ and the number of galaxies in a subcluster $N_g$. As can be seen from Figure 8a, higher mass fractions of dark matter do not always imply greater final segregations. Within the gravitational picture leading to equations (12) and (13) we see that, for $\eta$ values not very different from 0.9, higher values of this parameter imply a faster growth of the subcluster size and dynamical friction loses then effectiveness. Figure 9 shows the results of two simulations (D1 and D2) where different $\eta$ values were taken. We see that final segregation is in fact smaller when $\eta = 0.95$ than for $\eta = 0.8$. Simulations with smaller $\eta$ values have been performed. However in that case, a major part of mass is in galaxies, the initial system cannot be then considered as a uniform sphere. Consequently, equation (3) assignes a too large velocity to particles, and we do not observe a subsequent collapse of the system.

With respect to the expansion factor dependence, equation (12) implies slightly longer dynamical friction timescales for higher $R_m/R_{in}$ values. However, the collapse time is also longer and dynamical friction has then much more time to operate within the subclusters. Consequently, $t_{col}/T_{df}$ increases with the expansion factor and a greater segregation is expected (figure 8b). N-body simulations (D3 and D4) with $R_m/R_{in} = 3$ and 12 show in fact this behaviour (Figure 10).

Initial conditions in our numerical experiments specify the total number of galaxies instead of the number per subcluster entering in equation (12). Simulations D5 and D6 show however that for higher total $N_g$ values, merging of substructures begins at earlier evolutionary times and the substructures then contain a higher number of galaxies. Consequently, the final amount of segregation is smaller (Figure 11) than in simulation A1. These numerical results also suggest that the final segregation depends on the richness of the cluster (Serna & Alimi 1993).

# 6. Tidal Stripping of Galaxies with Massive Dark Halos.

As we have quoted in the introduction, inherent to all the above numerical experiments are the assumptions that galaxies interact elastically and dark matter is initially distributed as a smooth background. However, inelastic encounters of galaxies containing massive dark halos could lead to loss of energy by the stripped dark material (Eyles et al 1986). Non-luminous particles could then progressively concentrate in the core as they are removed from galaxies giving rise to final distributions where dark matter particles are more centrally condensed than galaxies. In order to analyze this possibility, we have performed two simulations (E1 and E2) where the inelasticity of halo encounters is taken into account. The approach used for these simulation is similar to that considered by Evrard (1986), but adapted to the present problem.

We have generated positions and velocities for 30 galaxies as in the previous simulations. Around each galaxy, we have then assigned a halo where positions for 272 dark matter particles are generated at random through the density profile

$$\rho(r) = \frac{M_h}{4\pi G r_t r^2} \qquad (14)$$

where $r$ is the radial distance to the galaxy center, $r_t$ is a truncation radius and $M_h = 272 m_{dm}$ is the halo mass. The value of the truncation radius has been taken rather extended in E1 ($r_t = \frac{3}{2}d_i$) and rather small in E2 ($r_t = \frac{1}{2}d_i$), where $d_i = R_{in}/N_g^{1/3}$ is the initial intergalactic spacing, in order to be compatible with the estimated size of dark halos from observations of galaxies in clusters (Mamon & Gerbal 1992). Peculiar velocities of dark matter particles in a given halo were assigned at random from a Gaussian distribution with one-dimensional velocity dispersion $\sigma = v_{scl}/(2)^{1/2}$, where $v_{scl}^2 = GM_h/r_t$ is a scale velocity.

Figure 12 shows a very similar evolution of particle distribution for the simulations E1 and E2. However it is very different from previous simulations, because it is modified by the introduction of tidal stripping. Dark halos are disrupted rapidly and subclusters resulting from merging of galaxy and dark matter remain isolated. Finally we do not observe as in the previous simulation a spherical central overdensity. The harmonic mean radius for dark matter (Figure 13) appears essentially constant. However, final segregation holds always. Galaxies are much more clustered than dark matter particles. Similar results were obtained in experience E2. The segregation seems then rather persistent for any hierarchical clustering scenario, even when very different initial conditions are considered.

## 7. Conclusions.

We have performed a series of N-body experiments in order to study the formation of galaxy clusters containing a dominant dark matter component. To this end, we have considered the simplest cosmological scenario where clusters are formed in a bottom-up manner from the spherical collapse of a primordial density fluctuation. We confirm the numerical results of previous authors who found final distributions with a considerable spatial segregation between the two mass components. We have shown that such segregation cannot be explained by some biasing effect resulting from the initial distribution of galaxies in the deepest potential wells.

The cluster evolution is characterized by the development of dark matter subcondensations which grow and then merge giving rise to a final configuration without subclusters. Our results strongly point out that dynamical friction inside such substructures is responsible for the luminous-dark matter segregation. The origin of subclustering is due to two kinds of local gravitational perturbations: those produced by statistical density fluctuations in the initial particle distribution and those produced by the presence itself of very massive objects (galaxies). It has been shown that the first type has only a small influence on the final segregation. Since galaxies are always present in a hierarchical scenario, the final amount of segregation should be only very slightly modified by considering different initial fluctuation spectra.

On the contrary, there are physical parameters characterizing the model of protocluster, which have an important influence on the final result. Our analytical and numerical results with values for $\eta$ not excessively different from 0.9, imply a $\eta$ dependence on the final amount of segregation which is contrary to that expected from prejudices a priori: for higher $\eta$ values, smaller segregation results. But, in any case, a considerable effect is always obtained for this range of $\eta$ values.

We have also found that the expansion factor has a strong influence on the final segregation, which is significantly reduced when small $R_m/R_{in}$ values are considered. In the extreme case, where $R_m/R_{in}$ is as small as unity (simulation C1), no final segregation is found. From equation (3) we can consider the expansion factor as a measure for the starting time, that is, for the redshift formation of the cluster. This then implies that, for a given present density parameter $\Omega_0$, the effect will be smaller for the youngest clusters. The extreme but unrealistic case where cluster formation occurs at very recent epochs ($R_m/R_{in} \simeq 1$ or $z_i \simeq 0$), should give no final segregation.

Finally, the spatial distribution of galaxies is similar to that of dark matter particles when the total number of galaxies is very high: The richer the cluster, the smaller is the luminous-dark matter segregation.

Tidal stripping between massive dark halos of galaxies modifies somewhat the dynamical evolution of clusters. However, final distributions for a large range of the halo radius show galaxies more clustered than dark matter particles.

Consequently, even though some particular young and very rich clusters could have little segregation, we conclude that this effect cannot be globally avoided in any hierarchical clustering scenario.

Observational evidences for such a segregation are rather uncertain giving contradictory results for or against such an effect. The dark matter density profiles obtained from X-ray observations show a tendency to be more centrally condensed than those of galaxies (Eyles et al. 1991; Buote & Canizares 1992; Gerbal et al. 1992; Watt et al. 1992), which is contrary to that predicted by the effect analyzed in this paper. In the opposite, there is an observed tendency of increasing the mass-to-light ratio with system size (Hoffman 1982 *et al.*) These last authors suggested that some segregation induced by dynamical friction could explain this observational tendency. Our numerical results agree in fact with those observations.

In any case, all the above observational results have their own set of assumptions, difficulties and uncertainties and none of them can be taken at present as a conclusive proof for or against such way of segregation, that is, for or against hierarchical clustering scenarios. However if such a segregation is present in real clusters, all observational analyses assuming that galaxies in clusters trace mass are misleading and their conclusions have to be reconsidered (Serna & Alimi 1993).

## Acknowledgments.

We thank D. Gerbal and M. J. West for useful discussions. We are specially indebted to S. D. M. White for several suggestions on this paper. A.S. thanks the Ministere de l'Education Nationale, France, for a post-doctoral fellowship. The computations were carried out on the CM-5 of the Institute du Physique du Globe (IPG), Paris.

# References.

# Figure Captions.

**Figure 1.**

Evolutionary stages for a typical 8192 particle system containing 30 galaxies (big points) and a non-luminous component (small points) of total mass fraction $\eta = 0.9$. The isodensity contours represent the dark matter distribution at intervals of 2.5 times the mean non-luminous density.

**Figure 2.**

Time evolution of the harmonic mean separation of dark matter particles (solid curve) and galaxies (dashed curve) for simulation A1.

**Figure 3.**

Evolutionary stages for a dark matter system.

**Figure 4.**

Time evolution of the harmonic mean separation for dark matter particles (solid curve) and for "galaxies" defined as those particles placed at the deepest potential wells (dashed curve). a) Both particle species have equal individual masses, b) The $m_g/m_{dm}$ ratio is similar to that of simulation A1..

**Figure 5.**

Time evolution of the harmonic mean separation of dark matter particles (solid curve) and galaxies (dashed curve) for a simulation where the expansion phase was canceled and particle positions were assigned at random.

**Figure 6.**

Evolutionary stages for a 8192 particle system containing 32 galaxies (big points) placed on a grid and dark matter particles (small points) placed on a different grid.

**Figure 7.**

Time evolution of the harmonic mean separation of dark matter particles (solid curve) and galaxies (dashed curve) for two different grid simulations with a) all particles placed on the same grid and b) each mass component on a different grid.

**Figure 8.**

Predicted $(t_{coll}/T_{df})$ at different times: $t_{coll}/4$ (solid curve), $t_{coll}/2$ (dashed curve) and $3t_{coll}/4$ (dashed-dotted curve) as a function of a) the non-luminous mass fraction of the overall cluster, b) the expansion factor and c) the number of galaxies in a subcluster.

**Figure 9.**

Time evolution of $R_{hm}$ for both components in clusters with different mass fractions of dark matter: a) $\eta = 0.8$, b) $\eta = 0.95$.

**Figure 10.**

Time evolution of $R_{hm}$ for both components in clusters with different expansion factors: a) $R_m/R_{in} \simeq 3$ and b) $R_m/R_{in} \simeq 12$.

**Figure 11.**

Time evolution of $R_{hm}$ for both components in clusters with different total numbers of galaxies: a) $N_g = 100$ and b) $N_g = 200$.

**Figure 12.**

Evolutionary stages for a 8192 particle system containing 30 galaxies with massive dark matter halos. The isodensity contours represent the dark matter distribution at intervals of 5 times the mean non-luminous density. We represent in the left side, the simulation E1 (initial large halos), in the right side, the simulation E2 (initial smaller halos).

**Figure 13.**

Time evolution of $R_{hm}$ for both components in clusters with galaxies containing massive dark halos interacting inelastically (Simulation E1).

TABLE 1

SUMMARY OF N-BODY EXPERIMENTS

| Run | Positions | $N_T$ | $N_g$ | $R_m/R_{in}$ | $\eta$ |
|---|---|---|---|---|---|
| A1 | Random | 8192 | 30 | 6 | 0.9 |
| B1 | Pot.Wells | 8192 | 28 | 6 | 1.0 |
| B2 | Pot.Wells | 8192 | 28 | 6 | 0.9 |
| C1 | Random | 8192 | 30 | 1 | 0.9 |
| C2 | Grid | 8192 | 30 | 6 | 0.9 |
| C3 | 2-Grids | 8192 | 32 | 6 | 0.9 |
| D1 | Random | 3630 | 30 | 6 | 0.8 |
| D2 | Random | 16384 | 30 | 6 | 0.95 |
| D3 | Random | 8192 | 30 | 3 | 0.9 |
| D4 | Random | 8192 | 30 | 12 | 0.9 |
| D5 | Random | 8192 | 100 | 6 | 0.9 |
| D6 | Random | 8192 | 200 | 6 | 0.9 |
| E1 | Large Halos | 8190 | 30 | 6 | 0.9 |
| E2 | Small Halos | 8190 | 30 | 6 | 0.9 |

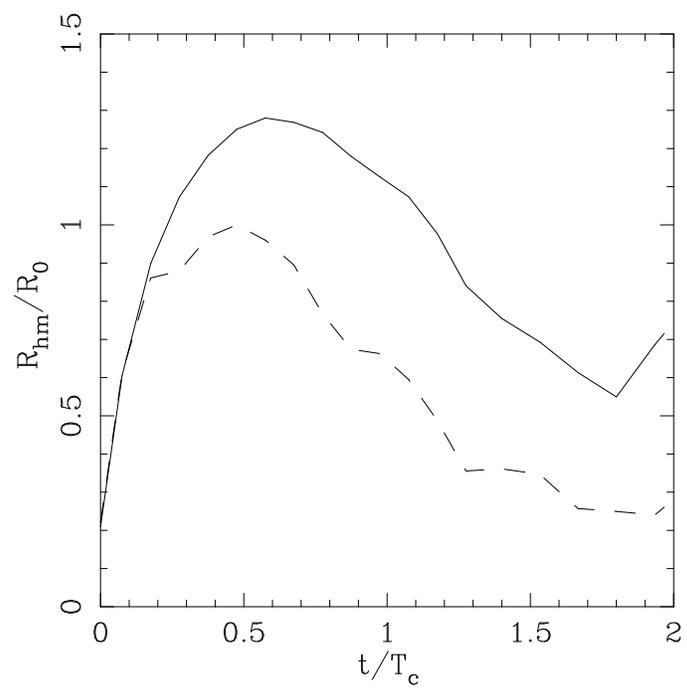

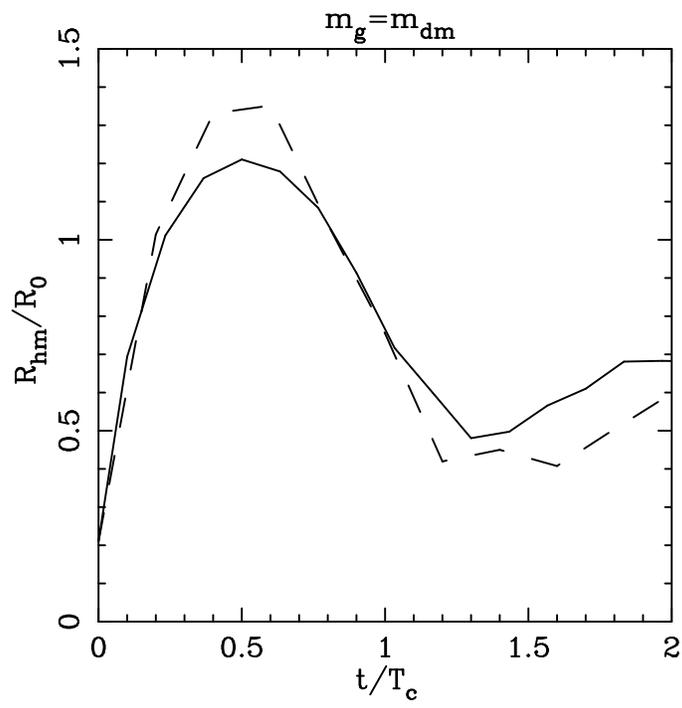

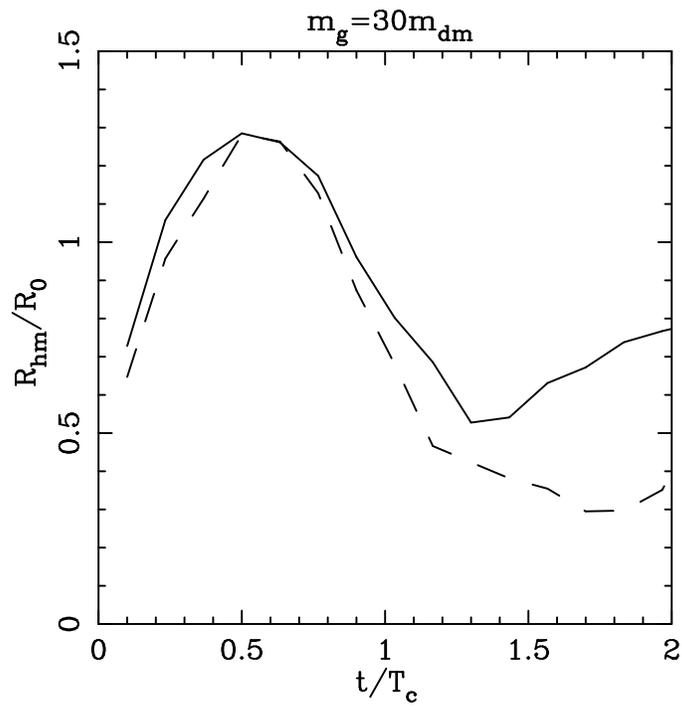

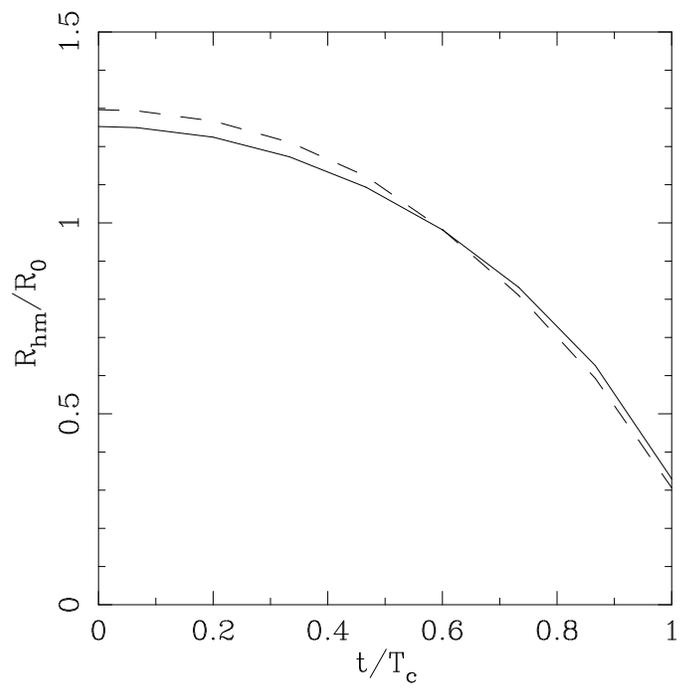

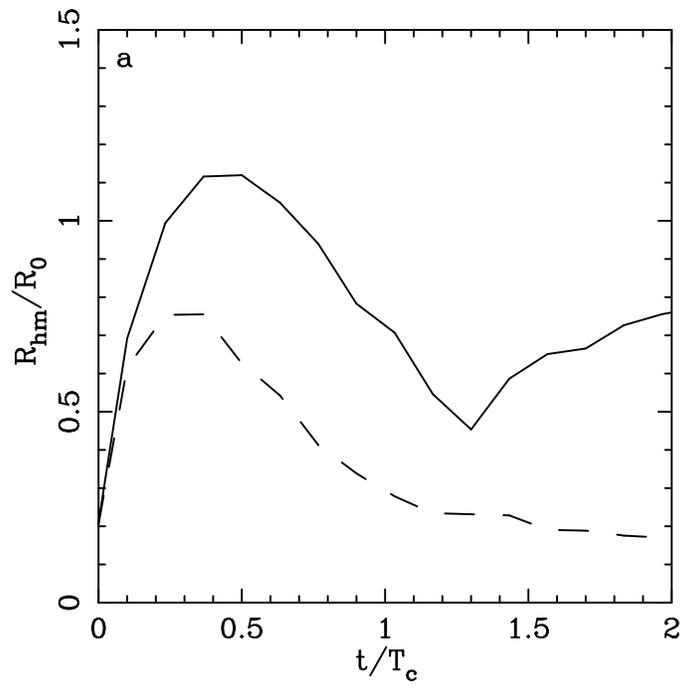
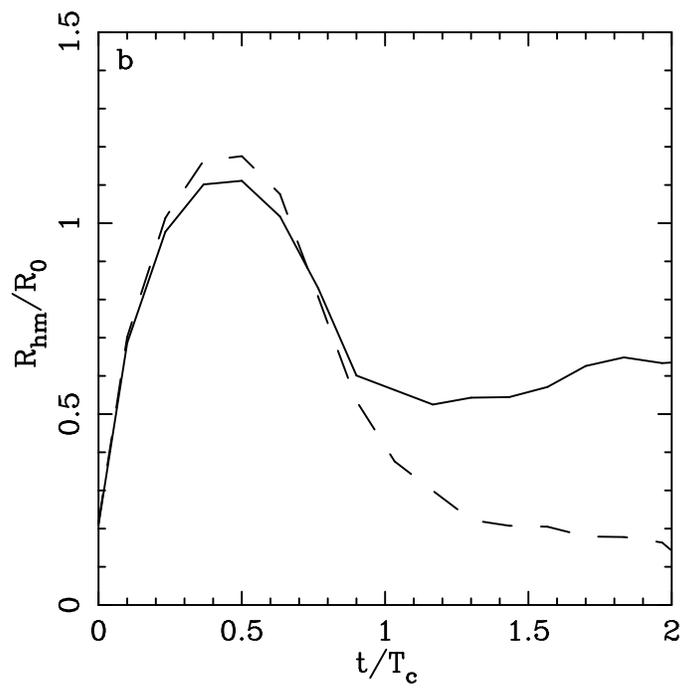

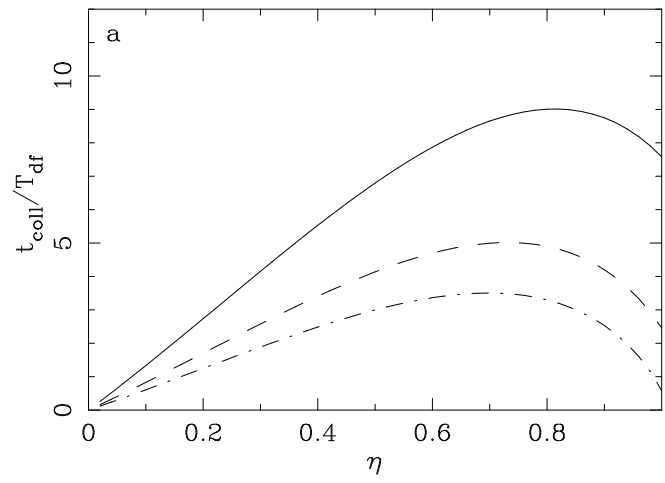
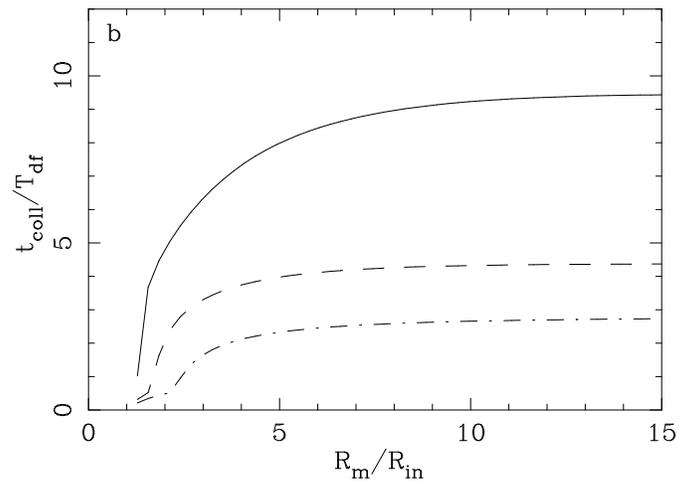
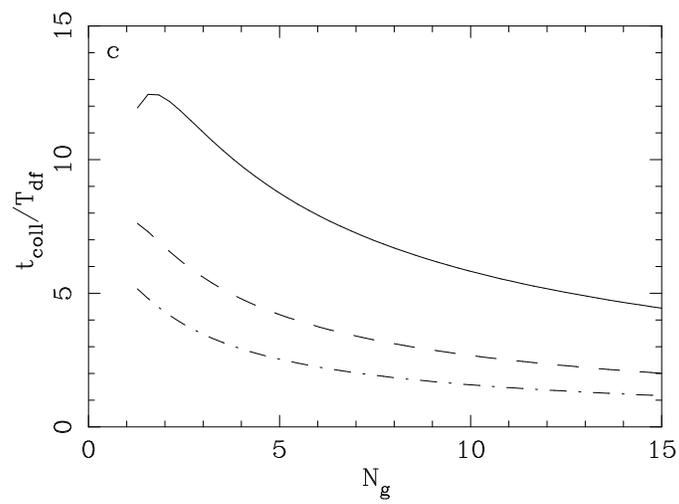

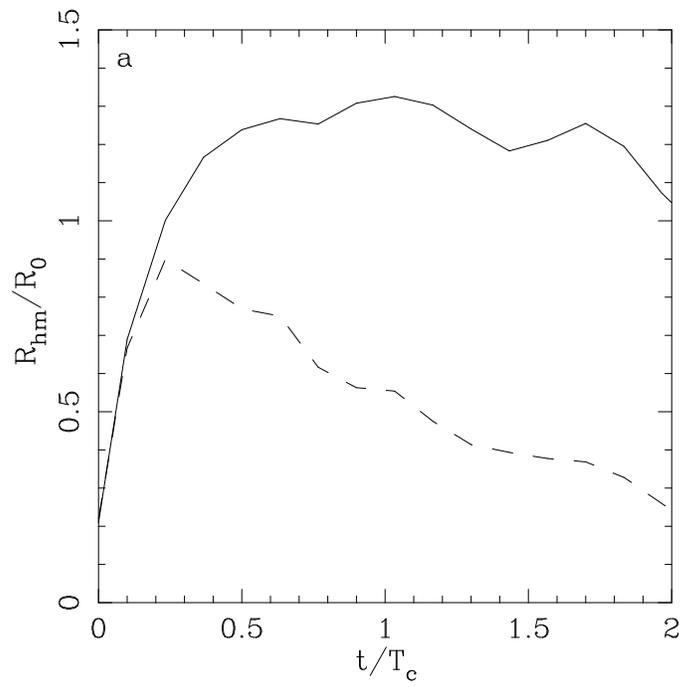

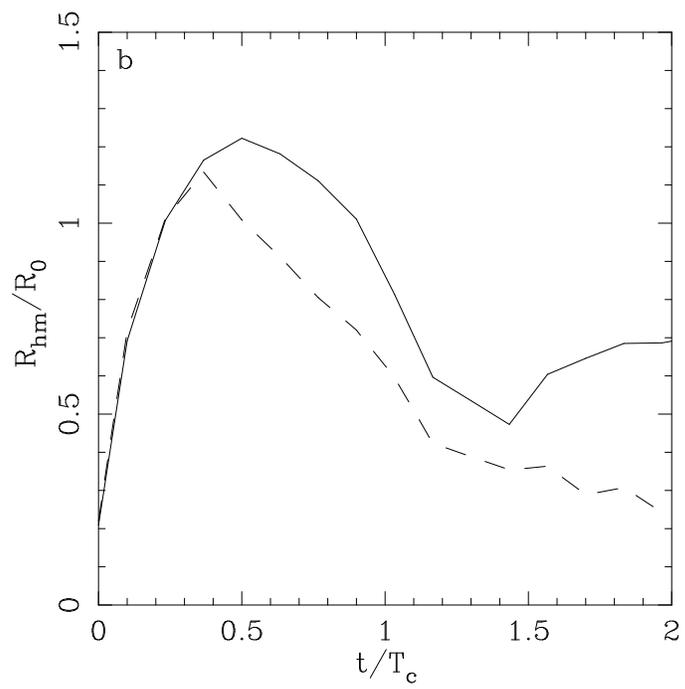

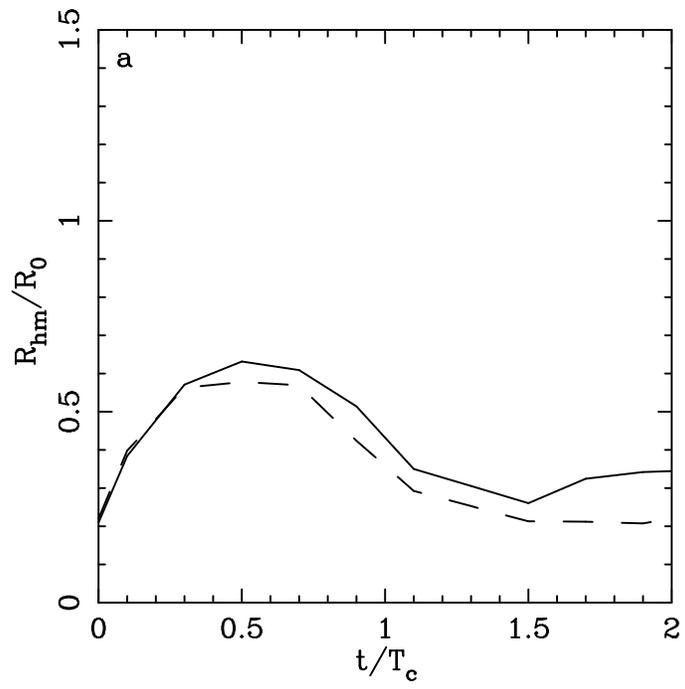

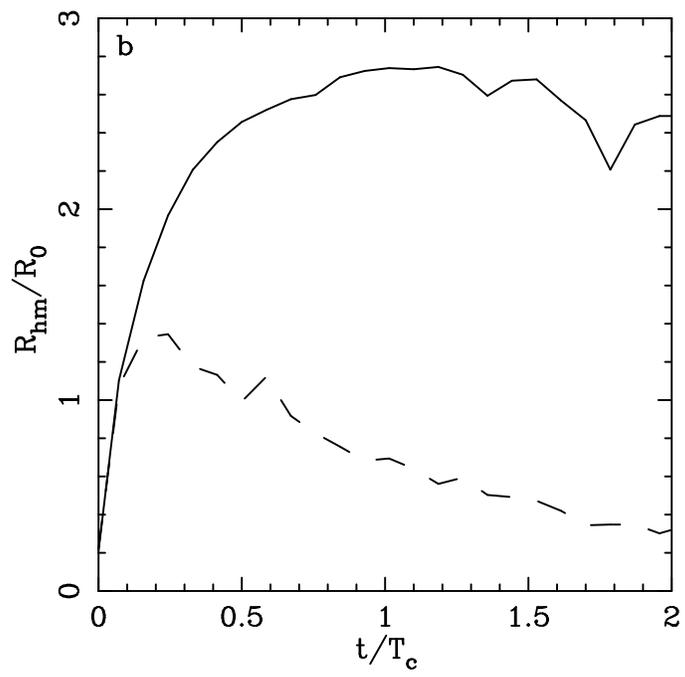

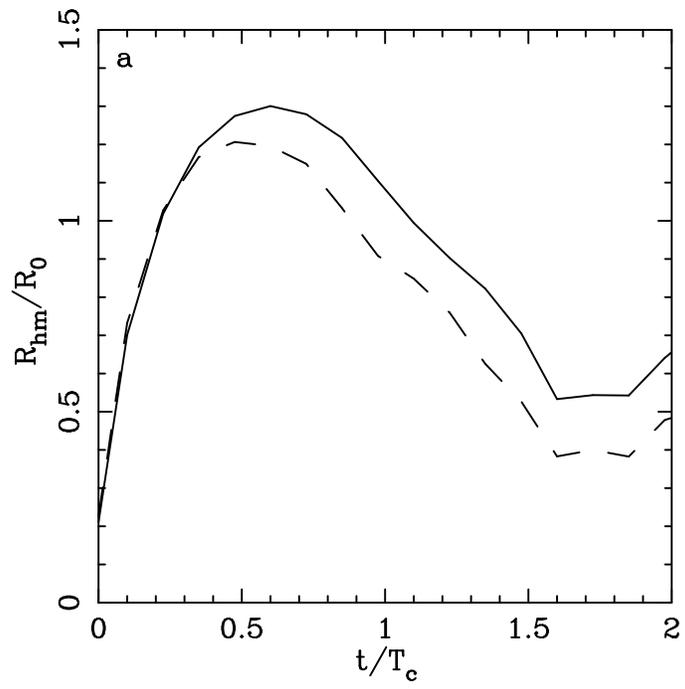

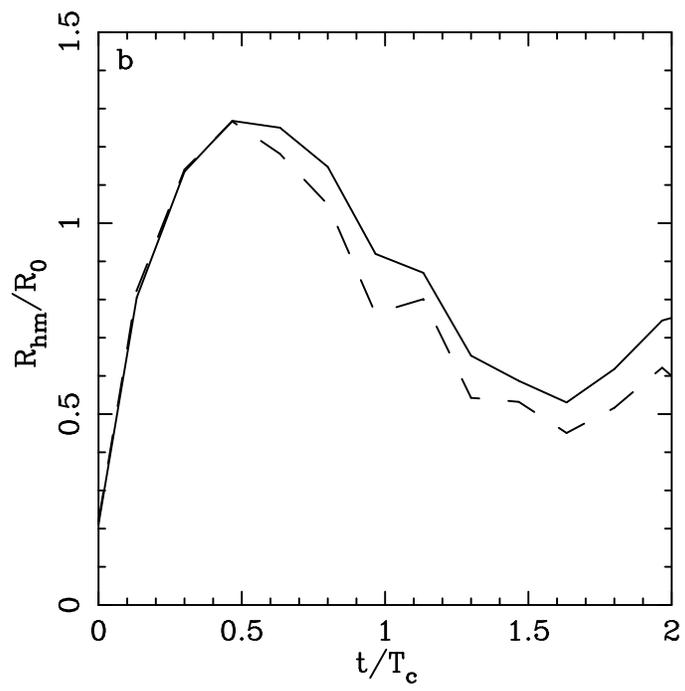

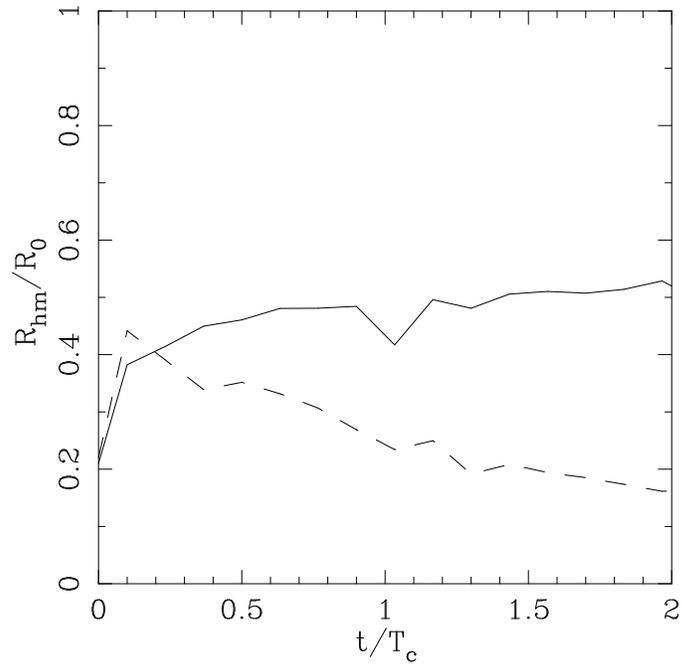